\documentclass[aps,pra,twocolumn,showpacs,floatfix]{revtex4}
\usepackage{epsfig}
\usepackage{graphicx}
\usepackage{dcolumn}
\usepackage{amsthm,amsmath}

\begin{document}

\title{Magnetic sublevel independent magic wavelengths: Application in the Rb and Cs atoms}
\author{Sukhjit Singh$^a$\footnote{Email: sukhjitphy.rsh@gndu.ac.in}, B. K. Sahoo$^{b}$\footnote{Email: bijaya@prl.res.in} 
and Bindiya Arora$^a$\footnote{Email: bindiya.phy@gndu.ac.in}}

\affiliation{$^a$Department of Physics, Guru Nanak Dev University, Amritsar, Punjab-143005, India}
\affiliation{$^b$Theoretical Physics Division, Physical Research Laboratory, Navrangpura, Ahmedabad-380009, India}
\date{Received date; Accepted date}
 
\begin{abstract}
A generic scheme to trap atoms at the magic wavelengths ($\lambda_{\rm{magic}}$s) that are independent of vector and tensor components 
of the interactions of the atoms with the external electric field is presented. The $\lambda_{\rm{magic}}$s for the laser cooling D2 lines in the Rb and 
Cs atoms are demonstrated and their corresponding polarizability values without vector and tensor contributions are given. Consequently, 
these $\lambda_{\rm{magic}}$s are independent of magnetic sublevels and hyperfine levels of the atomic states involved in the transition, thus, can offer unique approaches to carry out many high precision 
measurements with minimal systematics. Inevitably, the proposed technique can also be used for electronic or hyperfine transitions in other atomic systems.
\end{abstract}

\maketitle

{\it Introduction:} Techniques to cool and trap atoms using laser light have revolutionized modern experimental procedures. They are 
applied not only to carry out very high precision spectroscopy measurements, but also to probe many subtle signatures like parity 
violation \cite{wood}, Lorentz symmetry invariance \cite{haffner}, quantum phase transitions \cite{lukin} etc. Vogl and Weitz had 
demonstrated cooling of Rb atoms by resonating the trap laser light with their D-lines \cite{Vogl}, while Monroe et. al. had observed
the clock transition in Cs by cooing the atom using the D2 line ~\cite{monroe}.
As demonstrated in Ref. \cite{katori2}, trapping atoms at $\lambda_{\rm{magic}}$s is the foremost process today in a number of applications 
such as constructing optical lattice clocks. Following this, a number of experimental and theoretical studies have been reported 
$\lambda_{\rm{magic}}$s in the neutral atoms \cite{lundblad,marianna12,sahoo12,arora1,Takamoto,Yi,Barber,Ovsiannikov}, and recently in 
the singly charged alkaline earth ions \cite{Tang,jasmeet}. In a remarkable work, Katori et. al.~\cite{katori1} had demonstrated use of 
magic wavelengths ($\lambda_{\rm{magic}}$s) for Sr atoms, to reduce the systematics in the measurements. Use of $\lambda_{\rm{magic}}$s for trapping and controlling atoms inside high-Q 
cavities in the strong coupling regime with minimum decoherence for the D2 line of Cs atom have been demonstrated by McKeever et. 
al.~\cite{kimble1}.  
Liu and co-workers experimentally demonstrated the existence of $\lambda_{\rm{magic}}$s for the $^{40}$Ca$^+$ clock transitions~\cite{liu}.

A linearly polarized light is predominantly used to trap 
the atoms which is free from the contribution of vector component of the interaction between atomic states and electric fields. A substantial drawback of these $\lambda_{\rm{magic}}$s is that they are magnetic 
sublevel dependent for the transitions involving states with angular momentum greater than $1/2$. It has also been argued that considering 
circularly polarized light for trapping could be advantageous due to dominant role played by the vector polarizability in the ac 
Stark shifts \cite{sahoo12,us1}. This may help augmenting the number of $\lambda_{\rm{magic}}$s in some cases but at the same time requires magnetic 
sublevel selective trapping. The dependence of magic wavelengths on magnetic sublevels demands for state selective traps. To circumvent this problem, it is imperative to find out $\lambda_{\rm{magic}}$s independent of magnetic sublevel. 

 In this paper, we propose a scheme to trap atoms and ions at the $\lambda_{\rm{magic}}$s  that are independent 
of the atomic magnetic and hyperfine levels.  They can be used in a number of the applications discussed above. Just 
for the demonstration purpose, we present here $\lambda_{\rm{magic}}$s of the widely used D2 transitions of the Rb and Cs atoms. They are
useful for optical communications where lasers are tuned to their D-lines to trap and repump the atoms in order to prevent them from 
accumulating in the ground state \cite{Fox}. Moreover, D2 lines of Rb and Cs are used for studying their microwave spectroscopy 
\cite{Vogl,monroe,Yang,Entin}, quantum logic gates \cite{Friebel} and to assert the accuracy of the fine structure constant 
\cite{Gerginov}. In this proposal, we only presume that the atomic systems are trapped in sufficiently strong magnetic fields. 

\begin{figure}[t]
\centering 
\includegraphics[width=6.0cm, height=6.5cm]{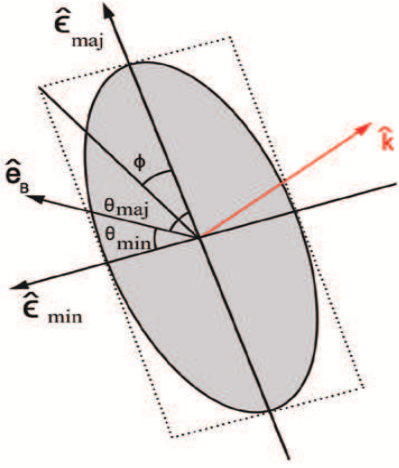}
\caption{(color online) A schematic representation of elliptically polarized light swept out by the polarization vector in one period. Unit vector 
$\hat{\epsilon}_{\rm{maj}}$ ($\hat{\epsilon}_{\rm{min}}$) aligns along the semi-major (semi-minor) axis. The vectors 
$\hat{\epsilon}_{\rm{maj}}$, $\hat{\epsilon}_{\rm{min}}$ and $\hat{k}$ are mutually perpendicular to each other, while $\hat{e}_B$ is the quantization 
axis lying in the plane of $\hat{\epsilon}_{\rm{maj}}$ and $\hat{\epsilon}_{\rm{min}}$ and is perpendicular to $\hat{k}$.}
\label{ellipse}
\end{figure} 

{\it Theory:} The ac Stark shift for any state with angular momentum $K$ of an atom placed in an oscillating electric field $\vec{\cal E}= \frac{1}{2}{\cal E} \hat{\epsilon} e^{-\iota \omega t
}+ c.c.$ with polarization vector $\hat{\epsilon}$ is given as~\cite{manakov}
\begin{equation}
\Delta E_K=-\frac{1}{2}\alpha_K(\omega){\cal E}^2,
\end{equation}
where $\alpha_K(\omega)$ is the total dynamic polarizability for the state $K$ with its magnetic projection $M$ as
\begin{eqnarray}
\alpha_K(\omega)&=&\alpha_K^{(0)}(\omega)+\beta(\epsilon)\frac{M}{2K}\alpha_K^{(1)}(\omega ) \nonumber \\ 
&& + \gamma(\epsilon) \frac{3M^2-K(K+1)}{K(2K-1)}\alpha_K^{(2)} (\omega),\label{eqmag} 
\end{eqnarray}
where $\alpha_K^{(i)}(\omega)$ with $i=0,1,2$ are the scalar, vector and tensor components of the frequency dependent polarizability respectively.In
the above expression $K$ and $M$ can be replaced suitably by either the atomic angular momentum $J$ or hyperfine angular momentum $F$ with their
corresponding magnetic projection $M_J$ or $M_F$, depending upon the consideration of atomic or hyperfine states, respectively. The $\beta(\epsilon)$ and 
$\gamma(\epsilon)$ are defined as~\cite{Beloy}
\begin{equation}
\beta(\epsilon) = \iota(\hat{\epsilon}\times\hat{\epsilon}^*)\cdot\hat{e}_B\label{beta}
\end{equation}
and
\begin{equation}
\gamma(\epsilon)= \frac{1}{2}[3(\hat{\epsilon}^*\cdot\hat{e}_B)(\hat{\epsilon}\cdot\hat{e}_B)-1], \label{gamma}
\end{equation}
with the quantization axis unit vector $\hat{e}_B$. The differential Stark shift of a transition between states $K$ to $K'$ is, hence, given by 
\begin{eqnarray}
\delta E_{KK'} &=& \Delta E_K - \Delta E_{K'} = -\frac{1}{2} \big [ \big\{\alpha_K^{(0)} ( \omega)-\alpha_{K'}^{(0)}(\omega) \big \} \nonumber \\ && 
+ \beta(\epsilon) \big \{ \frac{M}{2K}\alpha_K^{(1)}(\omega)  - \frac{M'}{2K'}\alpha_{K'}^{(1)}(\omega) \big \}  \nonumber \\ 
&& + \gamma(\epsilon) \big \{ \frac{3M^2-K(K+1)}{K(2K-1)}\alpha_K^{(2)} (\omega) \nonumber \\ 
&& -  \frac{3M'^2-K'(K'+1)}{K'(2K'-1)}\alpha_{K'}^{(2)} (\omega) \big \} \big ] {\cal E}^2, \label{eqmag1} 
\end{eqnarray}
It is obvious from the above expression that for obtaining null differential Stark shift, it is necessary that either independent components 
cancel out each other or the net contribution nullifies which prominently depends on the choices of $M$, $\beta(\epsilon)$ and $\gamma(\epsilon)$
values. By adequately selecting the experimental configuration such that the $\beta(\epsilon)$ and $\gamma(\epsilon)$ values are zero, it is 
possible to remove the vector and tensor components. As implied from above considerations and Eq. (5), the differential ac Stark shift depends 
only on the scalar polarizabilities of the associated states. Thus, it is independent of the magnetic sublevels. Moreover, the scalar 
polarizabilities are same for the atomic and hyperfine levels; i.e. $\alpha_J^{(0)}=\alpha_F^{(0)}$ (for all the allowed $F$ values). Hence,
these $\lambda_{\rm{magic}}$s are also independent of the hyperfine splittings of the participating atomic states. Therefore, 
$\lambda_{\rm{magic}}$s obtained by applying the above conditions will be independent of the choice of $M_J$, $F$ and $M_F$ quantum numbers. 
On account of above, we elucidate a lab frame in which null values for $\beta(\epsilon)$ and $\gamma(\epsilon)$ can be accomplished.  

\begin{figure}
\includegraphics[width=\columnwidth,keepaspectratio]{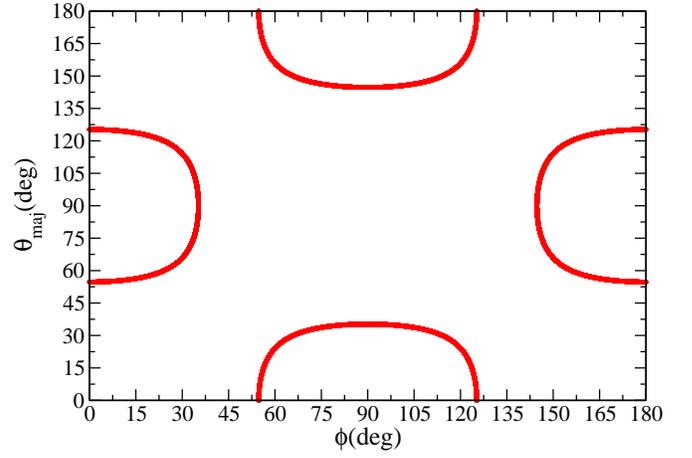}
\caption{(color online) Combinations of $\theta_{\rm{maj}}$ and $\phi$ values, marked with red curves, for which $\cos^2\theta_p=\frac{1}{3}$.}
\label{phitheta}
\end{figure}

{\it Discussions:} We start our analysis by defining a coordinate system with the components $\hat{\epsilon}_{\rm{maj}}$, $\hat{\epsilon}_{\rm{min}}$ 
and $\hat{k}$, where $\hat{\epsilon}_{\rm{maj}}$ and $\hat{\epsilon}_{\rm{min}}$ are the real components of the complex unit polarization vector 
$\hat{\epsilon}$ and describe the system such that 
\begin{equation}
\hat{\epsilon}=e^{\iota \sigma}(\cos\phi\hat{\epsilon}_{\rm{maj}}+\iota\sin\phi\hat{\epsilon}_{\rm{min}})\label{epsilon} .
\end{equation}
Here parameter $\phi$ is analogous to the degree of polarization ($A$) and $\sigma$ is a real quantity representing an arbitrary phase. Conveniently this 
can be represented by Fig. \ref{ellipse}, where the electric field vector sweeps out an ellipse in a unit period about the axis of wave vector $\hat{k}$ 
with semi-major and semi-minor axes of the ellipse aligned along $\hat{\epsilon}_{\rm{maj}}$ and $\hat{\epsilon}_{\rm{min}}$, respectively. The ratio of 
the semi-minor width to semi-major width of the ellipse needs to be $\tan\phi$. Furthermore using Eqs. (\ref{beta}) and (\ref{epsilon}), one can express 
$\iota (\hat{\epsilon}\times \hat{\epsilon}^*) = A\hat{k}$, where $A=\sin2\phi$. The biased magnetic field is along the quantization axis $\hat{e}_B$ and 
can technically be aligned in any direction. Without loss of generality, it can be assumed to lie in the $\hat{\epsilon}_{\rm{maj}} \sim
\hat{\epsilon}_{\rm{min}}$ plane for the present requirement. Parameters $\theta_{\rm{maj}}$, $\theta_{\rm{min}}$ and $\theta_{k}$ are the angles between 
the respective unit vectors and $\hat{e}_B$, respectively, as shown in Fig. \ref{ellipse}. In terms of these geometrical parameters, one can conveniently
express $\beta(\epsilon)=A\hat{k}.\hat{e}_B = A\cos\theta_k$ and $\gamma(\epsilon)=\frac{1}{2}(3\cos^2\theta_p-1)$ satisfying the relations
\begin{eqnarray}
\label{thetaeq}
\cos^2\theta_p=\cos^2\phi\cos^2\theta_{maj}+\sin^2\phi\cos^2\theta_{min} 
\end{eqnarray}
and
\begin{eqnarray}
\theta_{min}+\theta_{maj}=90^\circ .
\end{eqnarray}
Substituting explicit form of $\beta(\epsilon)$ and $\gamma(\epsilon)$ in Eq.(\ref{eqmag}), the expression for 
the polarizability is given by
\begin{eqnarray}
\alpha_K(\omega)&=&\alpha_K^{(0)}(\omega)+(A\cos\theta_k)\frac{M}{2K}\alpha_K^{(1)}(\omega )+ \nonumber \\ 
 && \left(\frac{3\cos^2\theta_p-1}{2}\right) \frac{3M^2-K(K+1)}{K(2K-1)}\alpha_K^{(2)} (\omega).
\end{eqnarray}
In this description, it reduces to $\beta(\epsilon)=0$ and $\gamma(\epsilon)=\frac{1}{2}(3\cos^2\psi-1)$ for the 
linearly polarized light with $\phi=0$, where $\psi$ is the angle between the quantization axis and direction of polarization vector.
Similarly, one can simplify the above expression for the circularly polarized light by using either $\phi=45^{\circ}$ or 
$\phi=135^{\circ}$.

\begin{figure}
\includegraphics[width=\columnwidth,keepaspectratio]{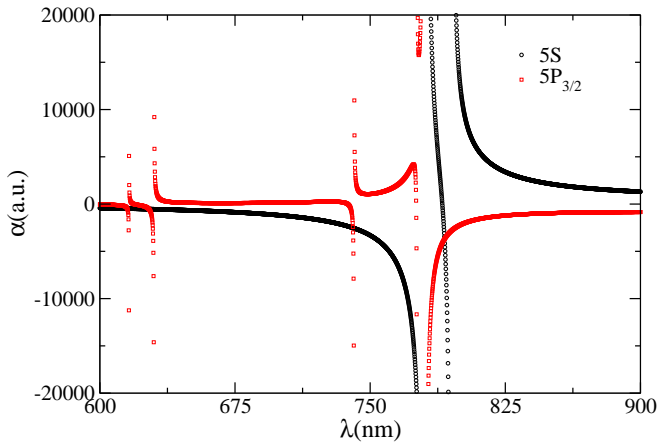}
\caption{(color online) $M_J$ independent dynamic polarizabilities (in a.u.) of the $5S$ and $5P_{3/2}$ states in Rb against wavelengths (in nm) for 
the proposed elliptically polarized light with $\cos\theta_k=0$ and $\cos^2\theta_p=\frac{1}{3}$.}
\label{freemj2}
\end{figure}

To eliminate the dependence of $\lambda_{\rm{magic}}$ on $M$ values in Eq. (\ref{eqmag1}), one can choose a suitable combination of the above parameters 
so that null values for both $\beta(\epsilon)= A\cos\theta_k$ and $\gamma(\epsilon)=\frac{1}{2}(3\cos^2\theta_p-1)$ can be achieved. This obviously corresponds to $\cos\theta_k=0$ and $\cos^2\theta_p=
\frac{1}{3}$, which can be brought about by suitably setting up the $\phi$, $\theta_k$ and $\theta_{\rm{maj}}$ parameters. One can achieve $\cos\theta_k=0$
by fixing the quantization axis $\hat{e}_B$ at right angle to the wave vector; i.e. one can assume $e_B$ to repose in the $\hat{\epsilon}_{\rm{maj}} \sim 
\hat{\epsilon}_{\rm{min}}$ plane. On the otherhand, many possible freedom exist to achieve $\cos^2\theta_p=\frac{1}{3}$. 
For example, we plot $\theta_{\rm{maj}}$ values (note that $\theta_{\rm{maj}}$ and $\theta_{\rm{min}}$ are related) versus $\phi$ in Fig. \ref{phitheta}, 
where each point on the graph represents a set of $\theta_{\rm{maj}}$ and $\phi$ that can yield $\cos^2\theta_p=\frac{1}{3}$. As mentioned 
previously, $\phi$ is a measure of the polarization and can be adjusted by setting the eccentricity `$e$' of the ellipse 
$\left(|A|=\frac{2\sqrt{1-e^2}}
{2-e^2}\right)$. It is evident from Fig. 2 that for the values $\phi =45^{\circ}$ and 135$^{\circ}$, none of the pair of 
angles $\phi$ and $\theta_{\rm{maj}}$ can offer $\cos^2\theta _p=\frac{1}{3}$. This means that the above criterion cannot be attained by applying 
the circularly polarized light. However as has been reported in \cite{Kotochigova}, this condition can be attained with $\phi$ = 0$^{\circ}$ and 
$\psi$ = 54.74$^{\circ}$ for a linearly polarized light. This critical 
condition seems to be too demanding and could be hard to achieve in an experimental set-up. On the otherhand, one could get a more
relaxed experimental conditions by using an elliptical polarized light as it offers more freedom to choose from a number 
of $\phi$ and $\theta_{\rm{maj}}$ combinations as shown in Fig. \ref{phitheta}. In fact, we observe that these estimated values of 
$\lambda_{\rm{magic}}$s will change maximum up to 1\% for the variation of $\theta_{\rm{maj}}$ by one degree when $\phi$ is 
constant and vice versa. Therefore, it seems to be  feasible to prepare a trap geometry using our proposed criteria for 
the elliptically polarized light aptly.

\begin{table*}[t]
\caption{\label{magic1}Magic wavelengths $\lambda _{\rm {magic}}$s (in nm) with the corresponding polarizability $\alpha_v(\omega)$ (in atomic units) 
for the $5S-5P_{3/2}$ and $6S-6P_{3/2}$ transitions in the Rb and Cs atoms, respectively, with the proposed elliptically polarized light with 
$\cos\theta_k=0$ and $\cos^2\theta_p=\frac{1}{3}$. In between resonant wavelengths $\lambda_{\rm{res}}$s (in nm) are also mentioned.} 
\begin{center}
\begin{tabular}{ccccccccc}
\hline \hline
  \multicolumn{4}{c}{Rb}&  & \multicolumn{4}{c}{Cs}\\ 
  \cline{1-4} \cline{6-9} \\
 Resonant transition  &  $\lambda_{\rm{res}}$ &$\lambda _{\rm {magic}}$&$\alpha_v(\omega)$&  & Resonant transition &  $\lambda_{\rm{res}}$  &$\lambda _{\rm {magic}}$&$\alpha_v(\omega)$ \\
 \hline
  & & & & & \\
$5P_{3/2}-8D_{3/2}$ &543.33 &&&&$6P_{3/2}-9D_{5/2}$ &584.68 &&\\ 
& &  615.2  &  $-478$ &&&& 602.9& $-340$  \\ 
$5P_{3/2}-8S_{1/2}$ & 616.13&&& & $6P_{3/2}-10S_{1/2}$&603.58  &&\\ 
&& 627.2  &  $-532$ &&&&614.8&-368    \\
$5P_{3/2}-6D_{5/2}$ & 630.01 &&&&$6P_{3/2}-8D_{5/2}$&621.48 &&\\
$5P_{3/2}-6D_{3/2}$ & 630.10& & &&&&621.9& $-384$  \\ 
& &  740.4   &  $-2515$  &&$6P_{3/2}-8D_{3/2}$&621.93&&\\ 
$5P_{3/2}-7S_{1/2}$ & 741.02 &&&&&& 657.7& $-502$ \\
&&  775.8  &  $-20048$&&$6P_{3/2}-9S_{1/2}$ &658.83 && \\
$5P_{3/2}-5D_{5/2}$ & 775.98 &&&&&& 685.9& $-628$ \\  
$5P_{3/2}-5D_{3/2}$ & 776.16 &&&&$6P_{3/2}-7D_{5/2}$ &697.52 &&\\
$5P_{3/2}-5S_{1/2}$ & 780.24 &&&&&& 698.5& $-697$ \\
&&  791.3  &  $-3681$&&$6P_{3/2}-7D_{3/2}$ &698.54 &&\\          
$5P_{3/2}-6S_{1/2}$ & 1366.87 &&&&&& 793.6& $-2094$ \\  
&&  1397.1  &  461&&$6P_{3/2}-8S_{1/2}$ &794.61 &&\\
$5P_{3/2}-4D_{5/2}$ & 1529.26 &&&&$6S_{1/2}-6P_{3/2}$ &852.35 && \\\\
&&  &&&&& 886.4& $-3736$ \\
&&&&&$6S_{1/2}-6P_{1/2}$ &894.59 &&\\
&&&&&$6P_{3/2}-6D_{5/2}$ &917.48 &&\\       
&&&&&&& 920.6 & 4131\\   
&&&&&$6P_{3/2}-6D_{3/2}$ &921.11 &&\\ 
&&&&&&& 936.2&2994  \\
&&&&&$6P_{3/2}-7S_{1/2}$ &1469.89 &&\\         
\hline \hline
\end{tabular}
\end{center}
\end{table*}
 
{\it Results:} It looks straightforward to achieve $\lambda_{\rm{magic}}$s for any atomic or hyperfine transitions in a given atomic system by 
maintaining the above geometry for trapping atoms provided that the differential polarizabilities of the considered transition nullifies within 
the resonance lines. We subsequently demonstrate below these $\lambda_{\rm{magic}}$s, specifically for the D2 lines of the Rb and Cs atoms.  

\textit{Rb atom:} In Fig. \ref{freemj2}, we have plotted scalar dipole polarizabilities of the $5S$ and $5P_{3/2}$ states of Rb with respect 
to wavelength of the external electric field. These
values were obtained in our previous work where we presented $\lambda_{\rm{magic}}$s for the D lines of Rb using the linearly and circularly polarized light \cite{sahoo12}. As can be seen from the figure, a number of $\lambda_{\rm{magic}}$s represented by the crossings of $5S$ and $5P_{3/2}$  
polarizabilities have been predicted for this transition and are presented in Table \ref{magic1} along with the resonance lines to highlight their 
locations. Two $\lambda _{\rm {magic}}$s are found at 615.2 nm and 627.2 nm, which belong to the visible region, while the other five 
$\lambda _{\rm {magic}}$s are located at 740.4 nm, 775.8 nm, 791.3 nm and 1397.1 nm. One more probable $\lambda _{\rm {magic}}$ in between the $5P_{3/2}-6D_{5/2}$ 
and $5P_{3/2}-6D_{3/2}$ resonance lines seems to exist, but we have not listed it in the above table due to inability to identify it distinctly. All the $\lambda _{\rm {magic}}$s mentioned in Table. \ref{magic1}, except the one at 1397.1 nm, support blue-detuned trapping 
scheme. We, however, recommend the use of $\lambda_{rm{magic}}$ at 791.3 nm for a blue-detuned and 1397.1 nm for a red-detuned trap for the experimental 
purposes, since these wavelengths are far from the resonant transitions. 

\begin{figure}
\includegraphics[width=\columnwidth,keepaspectratio]{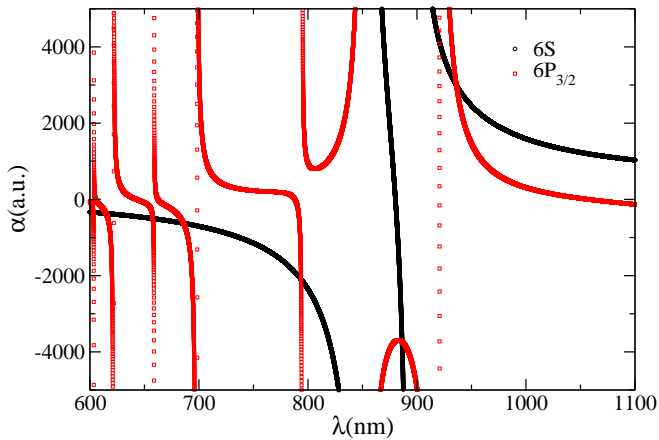}
\caption{(color online) $M_J$ independent dynamic polarizabilities (in a.u.) of the $6S$ and $6P_{3/2}$ states in Cs against wavelengths (in nm) for 
the proposed elliptically polarized light with $\cos\theta_k=0$ and $\cos^2\theta_p=\frac{1}{3}$.}
\label{freemj}
\end{figure} 

\textit{Cs atom:} Adopting a similar procedure as in Ref. \cite{arora1}, we have evaluated dynamic polarizabilities of the ground
and $6P$ states of Cs. We have also plotted the frequency dependent scalar polarizabilities of the ground and $6P_{3/2}$ states
of this atom in Fig. \ref{freemj} to find out $\lambda _{\rm {magic}}$s that are independent of the magnetic sublevels and hyperfine levels of the atomic 
states of the D2 line. Table \ref{magic1} enlists the $\lambda _{\rm {magic}}$s for this transition which lie within the wavelength range of 
600-1500 nm. As demonstrated in Ref.~\cite{arora1}, $\lambda _{\rm {magic}}$s exist between every two resonances for the linearly polarized 
light. We correspondingly locate six wavelengths at 602.9 nm, 614.8 nm, 621.9 nm, 657.7 nm, 685.9 nm and 
698.5 nm in the visible region using the proposed geometry for an elliptically polarized light. We also found two 
$\lambda _{\rm {magic}}$s at 793.6 nm and 886.4 nm, which belong to infra-red region. They all support dark or  blue-detuned traps. Two more $\lambda _{\rm 
{magic}}$s in the infrared region are located at 920.6 nm and 936.2 nm that can support red-detuned traps. In this case, we intend to recommend 
the use of $\lambda _{\rm {magic}}$s at 920.6 nm and 936.2 nm for the red-detuned trapping and 685.9 nm for the blue-detuned trapping due to the
availability of lasers at these wavelengths. 

{\it Conclusion}: A novel trap geometry has been proposed using an elliptically polarized light in a sufficiently large magnetic field that can produce
null differential Stark  shifts among the transitions and can be exclusively applicable among any magnetic sublevels and hyperfine levels. Their 
applications  in the D2 lines of Rb and Cs atoms have been highlighted and the corresponding magic wavelengths 
are reported. Furthermore, we have also recommended the magic wavelengths that are suitable for both the blue and red detuned traps of Rb and Cs atoms. 
These magic wavelengths will be immensely useful in a number of high precision measurements.

{\it Acknowledgement:} S.S. acknowledges financial support from UGC-BSR scheme. The work
of B.A. is supported by CSIR Grant No. 03(1268)/13/EMR-II, India. We
gratefully acknowledge helpful discussions with Dr. K. Beloy, Jasmeet Kaur and Kiranpreet Kaur.


\end{document}